\documentclass[twocolumn,showpacs,preprintnumbers,amsmath,amssymb]{revtex4}

\usepackage{graphicx}
\usepackage{dcolumn}
\usepackage{bm}

\begin{document}

\title{Anharmonic magnetic deformation of self-assembled
molecular nanocapsules}

\author{O.V. Manyuhina$^{1}$, I.O. Shklyarevskiy,$^{1,2}$ P. Jonkheijm,$^{2}$ P.C.M.
Christianen,$^{1}$ A. Fasolino,$^{1}$ M.I. Katsnelson,$^{1}$
A.P.H.J. Schenning,$^{2}$ E.W. Meijer,$^{2}$ O. Henze,$^{3}$ A.F.M.
Kilbinger,$^{3}$ W.J. Feast,$^{2,3}$ and J.C. Maan$^{1}$}

\address {
$^{1}$ Institute for Molecules and Materials, Radboud University
Nijmegen, Toernooiveld 7, 6525 ED Nijmegen, The Netherlands
         }
\address{
$^{2}$Laboratory of Macromolecular and Organic Chemistry,
Eindhoven University of Technology, P.O. Box 513, 5600 MB
Eindhoven, The Netherlands
        }
\address{
$^{3}$Department of Chemistry, University of Durham, Durham DH1 3LE,
U.K.        }

\date{\today}

\begin{abstract}
High magnetic fields were used to deform spherical nanocapsules,
self-assembled from bola-amphiphilic sexithiophene molecules. At low
fields the deformation -- measured through linear birefringence --
scales quadratically with the capsule radius and with the magnetic
field strength. These data confirm a long standing theoretical
prediction (W. Helfrich, Phys. Lett. {\bf 43A}, 409 (1973)), and
permits the determination of the bending rigidity of the capsules as
(2.6$\pm$0.8)$\times 10^{-21}$ J. At high fields, an enhanced
rigidity is found which cannot be explained within the Helfrich
model. We propose a complete form of the free energy functional that
accounts for this behaviour, and allows discussion of the formation
and stability of nanocapsules in solution.
\end{abstract}
\pacs{62.25.+g, 81.16.Fg, 82.70.Uv, 83.60.Np}

\maketitle

The spontaneous formation of hollow molecular spheres (vesicles) is
a prototype of self-assembly of organic molecules into well-defined
architectures. Vesicular self-assembly is a quite general
phenomenon~\cite{Antonietti} observed in lipid molecules (the
building blocks of cell membranes~\cite{Lasic}), polymers,
dendrimers, and $\pi$-conjugated oligomers~\cite{Hoeben}. This large
variety makes vesicles interesting for applications like tailor-made
nanocapsules for drug delivery~\cite{Antonietti},
opto-electronics~\cite{Piok}, and containers for chemical
reactions~\cite{Vriezema}. They also provide ideal systems to
develop physical models for self assembly. Such a model requires a
quantitative understanding of the weak non-covalent interactions
involved in the self-assembly process on the basis of which vesicle
formation and stability can be explained. To this end, we have
measured magnetic deformation of novel bola-amphiphilic~\cite{bola}
sexithiophene-based capsules~\cite{vesicle,Shklyarevskiy} and we
have developed a theory to describe this effect. Comparison between
theory and experiments allows the determination of important
quantities like the bending rigidity of the capsules and contribute
to an understanding of their stability.

The determination of the elasticity of vesicles through magnetic
deformation was proposed three decades ago by Helfrich
\cite{Helfrich1,Helfrich2}, but it was never demonstrated
experimentally. A magnetic field is a thermodynamic parameter that
exerts well-defined magnetic forces to deform all capsules in
solution, in a controlled and contact-free way. Furthermore, the
field-induced deformation can be quantified via a linear
birefringence measurement. The results are, therefore, easier to
compare with theory than those of other deformation methods, such as
using capillary forces \cite{Needham,Evans,Zhelev}, pressure
\cite{Zhong,Chong}, electric fields \cite{Kummrow,Peikov},
mechanical forces by an atomic force microscope tip \cite{Shao} or
optical tweezers \cite{Lee}.

Theoretically we use a fourth order Landau-Ginzburg expression of
the capsule free energy that describes quantitatively the measured
elastic deformation up to high magnetic fields, far beyond the
quadratic dependence predicted by Helfrich at low fields
\cite{Helfrich2}. From the data we obtain the free energy as a
function of temperature that explains the overall stability of
nanocapsules and predicts their dissolution at high temperature. The
quantitative determination of the free energy can be used to develop
and validate models for the non-covalent interactions responsible
for the self-assembly process.

Experimentally we investigate sexithiophene (6T)
(2,5'''''-(R-2-methyl-3,6,9,12,15-pentaoxahexadecyl ester), Fig.~1a)
which is a $\pi$-conjugated oligomer with a rigid apolar
sexithiophene block substituted at both ends by polar ethylene oxide
chains. 6T forms hollow spherical self-assemblies in a 2-propanol
solution (Fig. 1b, details can be found in \cite{Shklyarevskiy}).
Dynamic light scattering (DLS) reveals a narrow distribution of
capsule sizes, with an average radius $R(T)$ ($\pm$10\%) that
increases with temperature $T$ from 56~nm at 20~$^{\circ}$C to
126~nm at 60~$^{\circ}$C (Table \ref{tab:table1}). At 70~$^{\circ}$C
the spherical shape is lost, while at 80$^{\circ}$C the scattering
intensity is zero because the 6T molecules are molecularly dissolved
\cite{Shklyarevskiy}.

6T molecules have a strongly anisotropic diamagnetic susceptibility
$\chi$, which means that the sexithiophene core tends to orient
along the field \cite{Maretbook}. Consequently, capsules are
deformed from a sphere to an oblate ellipsoid, where more molecules
are parallel (top and bottom of the capsule) than perpendicular to
the field (around the equator, Fig. 1c), reducing the magnetic
energy at the expense of elastic energy. This magnetic deformation
is probed by measuring the field induced optical birefringence
$\Delta n$ in a resistive 20~T magnet using a laser (HeNe,
632.85~nm) modulation technique.
\begin{figure}
\includegraphics[width=0.9\linewidth]{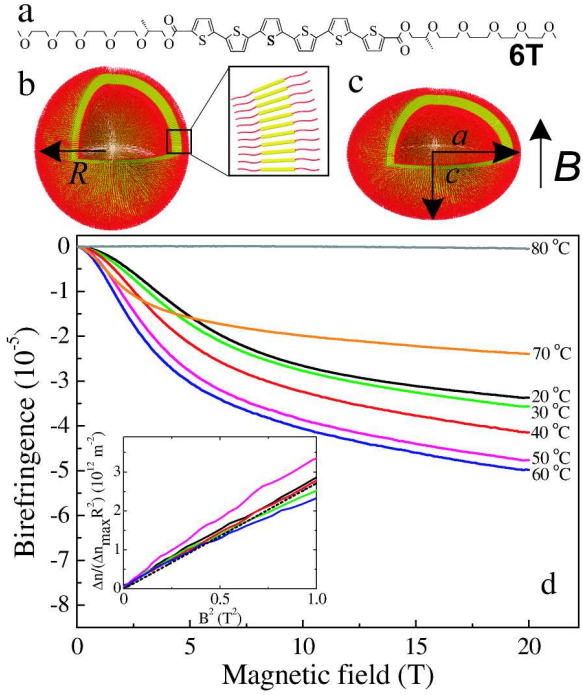}\\
\caption{\label{fig1} (color online) a) Chemical structure of a
sexithiophene molecule (6T). Schematic representation of a spherical
capsule with radius $R$ (b) and magnetically deformed capsule with
semi-axes $a$ and $c$ (c) d) Temperature dependence of the magnetic
field ($B$) induced birefringence of a 1~g/l 6T solution in
2-propanol. Inset: normalized birefringence Eq.(\ref{eq:birefr})
divided by $R^2$ and plotted versus $B^2$. The dashed line
corresponds to a fit with constant bending rigidity $k$.}
\end{figure}

Fig.~1d shows $\Delta n$ of a 1~g/l 6T/2-propanol solution at
different $T$. These curves were recorded during slow sweeps of the
magnetic field $B$ at constant $T$ ($\pm$0.1~$^{\circ}$C) and are
fully reproducible for up-, down-, and consecutive sweeps. In our
set-up a negative $\Delta n$ corresponds to 6T molecules aligning
with their long axis parallel to the magnetic field direction, as
expected for an oblate ellipsoid (Fig. 1c). The absolute value of
$\Delta n$ increases with $T$ up to 60~$^{\circ}$C. At
70~$^{\circ}$C it is considerably reduced with a different field
dependence, and $\Delta n$ is zero at 80~$^{\circ}$C, when the
capsules are dissolved. At low fields $\Delta n$ scales with the
square of the capsule radius and increases as $B^2$ with
approximately the same slope at all temperatures (inset of Fig.~1d).
These experimental facts prove that $\Delta n$ originates from a
reversible effect on stable nanocapsules below their melting
temperature (20~$^{\circ}$C$\div$60~$^{\circ}$C).

For small deformations (low fields) the normalized birefringence
$\Delta n$/$\Delta n_{\rm max}$ ($\Delta n_{\rm
max}=-6.5\cdot10^{-5}$ \cite{nmax}) varies linearly with
deformation~\cite{Helfrich1,Helfrich2,Maret}, but a more accurate
relation is needed for high deformations (high fields). For small
capsule concentrations $\rho_c$ the refractive index $n$ is related
to the polarizability of individual capsules $\alpha $ by
$n^{2}=n_{0}^{2}+4\pi \rho_c \alpha $, where $n_{0}$ is the
refractive index of the solvent. The dipole moment of the 6T
molecules $\mathbf{p}$ is connected with the effective electric
field $\mathbf{E}^{\prime }$ in the molecular layer by
$\mathbf{p}=\alpha _{\perp }\left( \mathbf{E}^{\prime
}\mathbf{-e}\left( \mathbf{eE}^{\prime }\right) \right) +\alpha
_{\parallel }\mathbf{e}\left( \mathbf{eE}^{\prime }\right)$, where
$\alpha _{\perp }$ and  $\alpha _{\parallel }$ are the
polarizabilities of 6T molecules with respect to $\mathbf{e,}$ the
unit vector normal to the capsule at a given point of its surface.
Applying standard boundary conditions on the Maxwell equations we
find the birefringence of a uniaxially deformed capsule:
\begin{equation} \label{eq:bi}
\Delta n=\pi \rho_c \rho_m \left( \alpha _{\parallel }\frac{\epsilon
_{0}}{\epsilon _{n}}-\alpha _{\perp }\right) \int \!dS\left(
1-3e_{z}^{2}\right),
\end{equation}
where $\epsilon _{0}$ and $\epsilon _{n}$ are the dielectric
constants of the solvent and of the 6T in the direction along
$\mathbf{e}$, respectively, $\rho_m$ is the surface density of
molecules in the layer, the integration is over the surface $S$ of
the capsule and $z$ labels the direction of magnetic field
$\mathbf{B}$.

\begin{figure}
\includegraphics[width=0.9\linewidth]{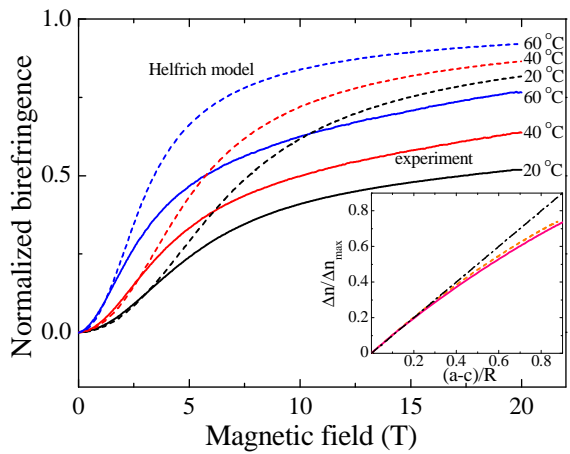}
\caption{\label{fig2} (color online) Normalized birefringence of
magnetically deformed 6T capsules. Solid lines: experimental data
for different temperatures. Dashed lines: calculated birefringence
in the Helfrich model, fitting the low-field behaviour using a
constant $k$ for all $T$. Inset: calculated relation between
birefringence and deformation for superspheroids (solid) and
spheroids (dashed) compared to $\Delta n/\Delta n_{\rm max}\simeq
(a-c)/R$ (dash-dotted line).}
\end{figure}

To calculate $\Delta n$ we assume that the shape of the deformed
capsules can be described by superspheroids, parameterized by :
\begin{equation}\label{eq:shape}
\begin{aligned}
x(u,v) &= a\cdot\sin^l v\cos u,\\
y(u,v) &= a\cdot\sin^l v\sin u,\\
z(u,v) &= c\cdot\cos^l v,\qquad (a>c),\\
\end{aligned}
\end{equation}
with $a$ is the radius in the circular $xy$ plane and $c$ the
extension in the $z$ direction (Fig.~\ref{fig1}c). Superspheroids
are more general than the spheroids ($l =1$) that are commonly used
for small deformations \cite{Helfrich1,Helfrich2}, and allow the
capsules to flatten ($l <1$). The birefringence now reads:
\begin{equation}\label{eq:birefr}
\frac{\Delta n}{\Delta n_{\rm max}} = -\frac {1}{2\pi a^2}\int
dS\,(1-3e_{z}^{2}).
\end{equation}
The inset of Fig.~\ref{fig2} shows $\Delta n$/$\Delta n_{\rm max}$
as function of deformation $(a-c)/R$ for (super)spheroids. We keep
the surface area of the capsule constant, assuming a constant number
of 6T molecules that form a permeable membrane. Remarkably, both
curves are rather similar, in particular in the low deformation
limit, where we recover the linear relationship between $\Delta n$
and deformation by expanding Eq.~(\ref{eq:birefr})
as~\cite{Helfrich1,Helfrich2,Maret}:
\begin{equation}\label{bir:expansion}
\frac{\Delta n}{\Delta n_{\rm max}}\simeq \frac{a-c}{R},
\end{equation}
which holds as long as $(a-c)/R \leq 0.3$ (inset Fig.~\ref{fig2}).

We determine the equilibrium values of $a,c$ and $l$ at each $B$ by
minimizing the total free energy ${\cal F_{\rm tot}}$, which is the
sum of the elastic ${\cal F_{\rm el}}$ and magnetic ${\cal F_{\rm
mag}}$ free energy. Simultaneously, the parameters in ${\cal F_{\rm
el}}$ are determined by a fit to the experimental $\Delta n$ using
relation~(\ref{eq:birefr}). We first consider the well known form of
the elastic term proposed by Helfrich~\cite{Helfrich2}, leading to:
\begin{equation}\label{eq:helfr}
\begin{gathered}
{\cal F_{\rm tot}} = {\cal F_{\rm el}} + {\cal F_{\rm mag}}+{\cal F_{\rm surf}}=
\frac k2 \int \!dS\, H^2
-\\
-\frac{\Delta \chi DB^2}{2\mu_{0}} \int \!dS \left (
e_{z}^{2}-\frac{1}{3}\right ) + \gamma\bigg(\!\int\! dS - 4\pi
R^2\bigg)^2\!,
\end{gathered}
\end{equation}
where $H=(\kappa_1+\kappa_2)/2$ is the mean curvature,
$\kappa_1,\kappa_2$ are the principal curvatures, $D=6.4$~nm (length
of long axis of 6T) is the membrane thickness, and $\Delta \chi =
\chi_{\perp}-\chi_{\parallel}$, the difference in magnetic
susceptibility along the short and long axes of 6T respectively
\cite{Sutter}. The term ${\cal F_{\rm surf}}$ imposes, for large
$\gamma$~\cite{gamma}, the constraint of constant capsule surface.
The bending rigidity $k$ is the only fit parameter to obtain the
best description of the experimental birefringence
(Fig.~\ref{fig2}). This approach can be used because the capsule
size is large compared to the intermolecular distances, which means
that the capsule shape is characterized by its curvatures.

\begin{table}
\caption{\label{tab:table1}Structural properties of 6T nanocapsules
in 2-propanol at different temperatures. Capsule radius $R$, as
determined by dynamic light scattering (exp.) and obtained by
fitting the magnetic birefringence (fit). Equilibrium parameters
$a$, $c$ and $l$ of the deformed capsules at 20~T.}
\begin{ruledtabular}
\begin{tabular}{cccccc}
Temperature&\multicolumn{2}{c}{Radius $R$ (nm)}&\multicolumn{3}{l}{Superspheroid(20 T)}\\
$T$ ($^{\circ}C$)& Exp.& Fit &$a/R$& $c/R$ &$l$\\
\hline
20& 56 & 58 &1.19 & 0.59 & 0.90\\
30& 68 & 62 &1.20 & 0.56 & 0.89\\
40& 75 & 75 &1.23& 0.50  & 0.88\\
50& 88 & 99 &1.26 & 0.42 & 0.85\\
60& 126 & 110 &1.28& 0.39 & 0.84\\
\end{tabular} \end{ruledtabular} \end{table}
Considering first the limit of small deformations we can expand
Eq.(\ref{eq:helfr}) as:
\begin{equation}\label{eq:helfr_K}
\frac{a-c}R = B^2\frac{R^2D\Delta \chi}{3 k\mu_0}.
\end{equation}
Combination of Eqs. (\ref{bir:expansion}) and (\ref{eq:helfr_K})
shows that the low field birefringence is inversely proportional to
the bending rigidity $k$, and scales quadratically with magnetic
field and capsule radius (the Helfrich result). Our experimental
results (Fig.~\ref{fig2}, inset of Fig.\ref{fig1}) indeed exhibit
this scaling up to $B\sim 1$ T, allowing to determine $k$ as
(2.6$\pm$0.8)$\times 10^{-21}$ J, independent of temperature. Our
data, therefore, permits the experimental determination of the
bending rigidity of this type of capsules.

The Helfrich free energy thus quantitatively describes the
birefringence data in the low deformation regime using a constant
bending rigidity. However, this model clearly overestimates the
experimental birefringence at high fields (Fig.~\ref{fig2}), which
indicates an increased bending rigidity at higher deformations.
Therefore we extend the expression of the free energy by including
all symmetry allowed terms up to fourth order \cite{Katsnelson}:
\begin{equation}\label{eq:complete}
{\cal F_{\rm el}}=\int \!dS [A'K+AH^2+c_1H^4+c_2H^2K+c_3K^2],
\end{equation}
where $K=\kappa_1\kappa_2$ is the Gaussian curvature. Note that
terms in $HK$ and $H^3$ are not allowed for symmetric
bola-amphiphiles~\cite{Katsnelson}. Since at $B=0$ the capsules are
spherical, with $H^2-K=0$, it is convenient to rewrite ${\cal F_{\rm
el}}$~as
\begin{multline}\label{eq:ham}
{\cal F_{\rm el}} =\int dS \big[(A+A')K+A(H^2-K) + (c_1+c_2+c_3)H^4-\\
         -(c_2+2c_3) H^2(H^2-K)+c_3(H^2-K)^2\big].
\end{multline}
If we compare the elastic energies Eq.(\ref{eq:ham}) for a flat
layer and for a set of spheres with radius $R$ with the same total
area, and neglecting the edge energy due to the interaction of the
solvent with the hydrophobic cores, we find that $A+A'<0$ and
$c_1+c_2+c_3>0$ must hold to favour the formation of spheres. In
this case, the minimum of Eq.(\ref{eq:ham}) occurs for a sphere of
radius:
\begin{equation}\label{eq:R_eq}
R^2=\frac{1}{H^2}=-2\frac{c_1+c_2+c_3}{A+A'}.
\end{equation}
Since the term $\int \!dSA'K$ does not depend on deformation, the
magnetic deformation data cannot provide directly information about
$A'$. In fact, according to the Gauss-Bonnet theorem, $\int \!dS K =
4\pi$ for any surface topologically equivalent to the sphere (like
the superspheroid considered here). However, Eq.(\ref{eq:R_eq}) can
be used a posteriori to determine $A'$, once the other parameters
are found by a best fit to the experiments. This approach is very
powerful, because it experimentally determines the total free energy
of nanocapsules, which is very difficult to obtain otherwise, and
which can be used to discuss the overall stability of capsules.

\begin{figure}
\includegraphics[width=0.99\linewidth]{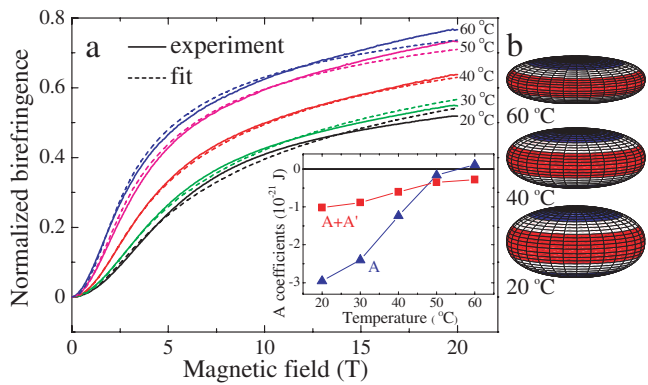}
 \caption{\label{fig3} (color online) a) Fit (dashed lines) of the experimental birefringence (solid
 lines) at different temperatures,
using superspheroidal capsules with deformation energy ${\cal F_{\rm
el}}$ given by equation~(\ref{eq:ham}). b) Equilibrium shapes of the
deformed capsules at 20~T.}
\end{figure}

Within this description the low field bending constant $k$ can be
calculated analytically for spheroids by expanding
Eq.(\ref{eq:complete}) for small deformations:
\begin{equation}\label{eq:complete_K}
k = 2A + \frac{28c_1+22c_2 + 16c_3}{3R^2},
\end{equation}
which connects our approach to that of Helfrich. Moreover, since $k$
has to be positive to guarantee stability against deformations, this
equation poses an extra constraint on the choice of parameters. In
the spirit of Landau theory, we assume that only $A$ depends on $T$
and that the coefficients $c_i$ are temperature independent. In
fact, by using the experimental value of $k$ we eliminate the
parameter $A$ from the fitting procedure, by which $c_i$ and $R(T)$
are determined via fitting to the experiments at the different
temperatures (Fig.~\ref{fig3}a).

For all temperatures the overall behavior of $\Delta n$ is nicely
reproduced, including the weaker field dependence at higher fields,
indicating anharmonic deformation characterized by a field enhanced
bending rigidity. The fitted value of $R(T)$ (Table
\ref{tab:table1}) is well within the error margins of the
experimental DLS values. Table \ref{tab:table1} also shows the
parameters $a$, $c$ and $l$ at 20 T, which correspond to
substantially deformed capsules, as illustrated in Fig.~\ref{fig3}b.
At the highest fields deformations $(a-c)/R$ as large as 0.9 are
found, which agrees very well with the observed deformation of 6T
capsules that are fixed in a compatible organogel and imaged by
scanning electron microscopy~\cite{Shklyarevskiy}. Not surprisingly,
we find that with increasing deformation the capsule shape becomes
flatter (decreasing $l$, see Table \ref{tab:table1}), maximizing the
number of molecules aligned along the field.

The best fits in Fig.~\ref{fig3}a are obtained using $c_1 = 1.148,
c_2 = -1.336, c_3 = 0.358 ~(10^{-35}~{\rm Jm}^2)$, and the
$T$-dependent $A$ and $A+A'$ are shown in the inset. Indeed, these
values make the capsules stable at $B=0$ ($A+A'<0, c_1+c_2+c_3>0$)
for $T<60^{\circ}C$. However, $A+A'$ approaches 0 when $T\to
60^{\circ}C$, suggesting an instability of the capsules at higher
$T$, which is indeed compatible with the experiments that show that
the capsules melt above $60^{\circ}C$. It is this vicinity of the
system to an instability, which makes fourth order terms comparable
in magnitude to second order ones and make the Helfrich model
inadequate for this situation.

The parameters $c_1,c_2,c_3$, $A(T)$ and $A'(T)$ fully describe the
free energy of the nanocapsules. Their actual values are intricately
related to the intermolecular interactions within the
capsule/solvent system, such as $\pi$-$\pi$ interactions between
neighboring 6T cores and interactions between 6T tails and solvent
molecules~\cite{Katsnelson, Jonkheijm}. Determination of these
parameters for a physical nanostructure therefore provides a
stringent test case, on the basis of which microscopic
intermolecular interaction models can be developed. Such a
quantitative description of non-covalent interactions within
molecular assemblies is an important step towards the detailed
understanding of the rules of self-assembly.

This work is part of the research programme of the 'Stichting voor
Fundamenteel Onderzoek der Materie (FOM)', which is financially
supported by the 'Nederlandse Organisatie voor Wetenschappelijk
Onderzoek (NWO)'.


\begin{thebibliography}{99}
\bibitem{vesicle} We use the term capsule instead of vesicle, which is usually used for lipid bilayer
systems in aqueous solutions.
\end{thebibliography}
\end{document}